# Direct Observation of a Uniaxial Stress-driven Lifshitz Transition in $Sr_2RuO_4$

V. Sunko[1,2], E. Abarca Morales[1,2], I. Marković[1,2], M.E. Barber[1], D. Milosavljević[1], F. Mazzola[2], D.A. Sokolov[1], N. Kikugawa[3], C. Cacho[4], P. Dudin[4], H. Rosner[1], C.W. Hicks[1,^], P.D.C. King[2,^] and A.P. Mackenzie[1,2,^]

[1] Max Planck Institute for Chemical Physics of Solids, Nöthnitzer Strasse 40, 01187 Dresden, Germany
[2] SUPA, School of Physics and Astronomy, University of St. Andrews, St. Andrews KY16 9SS, United Kingdom
[3] National Institute for Materials Science, Tsukuba, Ibaraki 305-0003, Japan.
[4] Diamond Light Source, Harwell Campus, Didcot, OX11 0DE, United Kingdom.
[^] to whom correspondence should be addressed: Clifford.Hicks@cpfs.mpg.de; philip.king@st-andrews.ac.uk; Andy.Mackenzie@cpfs.mpg.de

**Pressure represents a clean tuning parameter for traversing the complex phase diagrams of interacting electron systems[1,2], and as such has proved of key importance in the study of quantum materials. Application of controlled uniaxial pressure has recently been shown to more than double the transition temperature of the unconventional superconductor $Sr_2RuO_4$ for example[3–5], leading to a pronounced peak in $T_c$ vs. strain whose origin is still under active debate[4,6,7]. Here, we develop a simple and compact method to apply large uniaxial pressures passively in restricted sample environments, and utilize this to study the evolution of the electronic structure of $Sr_2RuO_4$ using angle-resolved photoemission. We directly visualize how uniaxial stress drives a Lifshitz transition of the γ-band Fermi surface, pointing to the key role of strain-tuning its associated van Hove singularity to the Fermi level in mediating the peak in $T_c$[7]. Our measurements provide stringent constraints for theoretical models of the strain-tuned electronic structure evolution of $Sr_2RuO_4$. More generally, our novel experimental approach opens the door to future studies of strain-tuned phase transitions not only using photoemission, but also other experimental techniques where large pressure cells or piezoelectric-based devices may be difficult to implement.**

The layered perovskite $Sr_2RuO_4$ has been extensively studied both because of its celebrated unconventional superconductivity [5,8–11] and the accuracy with which its normal state properties can be measured [12–15] and analysed [16–19]. In spite of a quarter of a century of work, there is still no consensus on the symmetry of its superconducting order parameter, or the mechanism by which the superconductivity condenses [5]. This is a major unsolved problem because its electronic structure, which is relatively simple compared to that of many other unconventional superconductors, is now known in considerable detail and its metallic state is firmly established to be a Fermi liquid below approximately 30 K [13]. A full understanding of the $Sr_2RuO_4$ problem is therefore a benchmark for the progress of the fields of strongly interacting systems and unconventional superconductivity.

Recent years have seen the development of uniaxial pressure as a new probe of the physics of $Sr_2RuO_4$ [3,4,7,20]. Unlike most unconventional superconductors, $Sr_2RuO_4$ has a long superconducting coherence length of over 70 nm, rendering the superconducting condensate the most sensitive to disorder of any known superconductor; the mean free path must be approximately 1 micron or larger for the superconductivity to be studied in the clean limit [21]. Any external tuning of the superconducting state must therefore preserve this extremely long mean free path, a constraint that has led to the failure of attempts to study the superconductivity while tuning the density of states at the Fermi level by chemical doping [22] or the application of biaxial epitaxial strain to thin films [23]. These issues can be overcome by the application of uniaxial pressure to high purity single crystals [3,4,7], which has been shown to raise $T_c$ from 1.5 K to 3.5 K. This observation explains the inhomogeneous traces of 3 K superconductivity that have been observed under externally imposed strain inhomogeneity[24], and around Ru inclusions in eutectic Ru-$Sr_2RuO_4$ mixtures [25].

Previous spectroscopic work[22,23] has identified the position of a van Hove singularity of the so-called γ band in unstrained $Sr_2RuO_4$ to be 14 meV above the Fermi level. Based on density-functional calculations [4], a working hypothesis has been that uniaxial pressure is driving the γ Fermi surface sheet through a Lifshitz transition.[4,7,26] Once traversed, the γ sheet would become open, a very unusual situation in an unconventional superconductor. However, it remains unclear if the intuitions based on single-particle calculations really represent a good starting point for considering strain-dependent changes to a Hund's metal system where orbital-dependent correlations are known to be highly important.[15,27] Indeed, it has been predicted that strain may alternatively trigger an intervening phase, such as a spin-density wave, which cuts off an increase in $T_c$ before the Lifshitz transition is reached [6]. It is thus crucial to obtain direct, $k$-resolved spectroscopic evidence for the electronic structure evolution that is taking place over a comparable strain range to that within which $T_c$ is known to peak.

In principle, angle-resolved photoemission spectroscopy (ARPES) is an ideal tool for this purpose, but this kind of experiment presents severe experimental challenges. Using piezoelectric-driven uniaxial pressure cells, as in [3,4,7,20], would require major re-engineering of conventional ARPES manipulators and careful shielding. For compatibility with present facilities, our goal here was to develop a sample stage that fits onto standard sample carriers, implying maximum dimensions of ca. 12 x 12 x 3 $mm^3$. Additionally, to study single crystals it must be possible to cleave samples mounted on the apparatus. For ARPES measurements, large strains have been applied to low-elastic-modulus materials [28], and spring- and piezo-based devices have been used for detwinning [29–32]. In Ref. [33] a bending mechanism was employed to apply large adjustments to the strain of a sample placed under strong uniaxial compression by its unusual thermal contraction. However, it has proved difficult to realise large strains in high-elastic-modulus materials in a general way. Indeed, in our first attempt using a spring-based rig driven by an adjustment screw actuated *ex-situ* (described in the Supplementary Information, Fig. S1), we could not achieve a uniaxial stress in $Sr_2RuO_4$ larger than its room-temperature elastic limit of ~0.2 GPa, well below the value required to reach the peak in $T_c$ [34]. In this paper, we report a new experimental design that uses differential thermal contraction to apply uniaxial stress gradually as the sample is cooled, and use it to successfully obtain ARPES data on $Sr_2RuO_4$ driven across its Lifshitz transition. In doing so we clarify the physics of this important correlated metal and superconductor, and demonstrate a technology that we believe will prove extremely useful for the study of a wide range of other materials.

Our custom strain rig is illustrated in Fig. 1a. Details of its design and operation are given in Methods; here we state the key point which is that differential thermal contraction of the Ti and Al support blocks delivers, upon cooling from room temperature to below ~40 K, a uniaxial compression of 0.6% (see Methods) to a sample platform in which there is excellent strain field homogeneity. Taking into account the Poisson's ratio of titanium, this yields an anisotropic strain $\varepsilon_{xx}$-$\varepsilon_{yy}$ of -0.8%, where $\varepsilon_{xx}$ is the longitudinal strain in the platform, $\varepsilon_{yy}$ the transverse strain, and negative values denote compression. We have confirmed that such an anisotropic strain is achieved through comparison of optical micrographs measured at room temperature and ~10K (Supplemental Fig. S2). The whole assembly fits comfortably on a standard flag-style sample plate (Fig. 1b), of the form commonly found in ultra-high vacuum based techniques such as ARPES or scanning probe methods. The sample, mounted on top of the platform, remains fully accessible for e.g. sample cleaving and subsequent measurement.

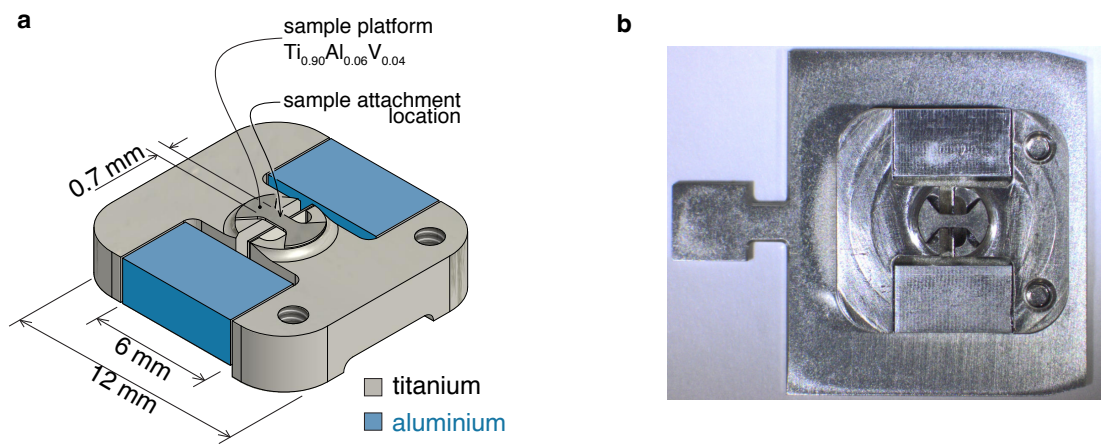

**Fig. 1: Differential thermal contraction strain rig.** (a) An illustration of the strain rig. The thermal contraction of aluminium exceeds that of titanium, leading to uniaxial compression of the sample platform during cooling. There is a copy of this platform on the underside, to maintain symmetry and avoid bending under the thermal stresses. Different parts of the device are joined by Stycast 2850. (b) A photograph of the strain rig mounted on a standard flag-style sample plate.

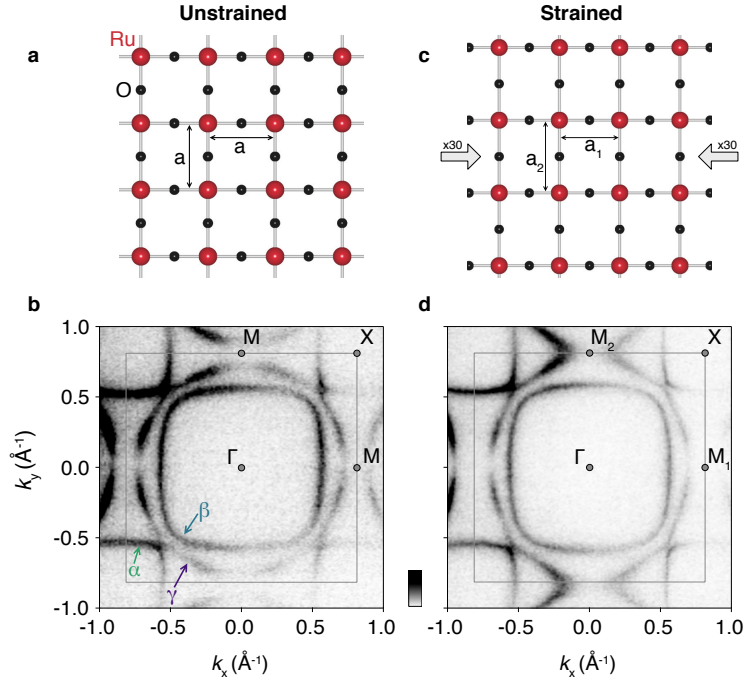

**Fig. 2: Strain-driven Lifshitz transition in Sr$_2$RuO$_4$.** (a) Schematic of the RuO$_2$ plane of Sr$_2$RuO$_4$, in its unstrained tetragonal phase. (b) ARPES measurements of the corresponding Fermi surface show the expected C$_4$ symmetry, with clear observation of the square hole pockets (α) located at the Brillouin zone corners and large nearly square (β) and circular (γ) electron-like pockets located at the zone centre. (c) Exaggerated (by a factor of 30) distortion of the RuO$_2$ plane by application of a uniaxial stress along [100], leading to an anisotropic strain of 0.7%. (d) A large distortion of the γ Fermi surface is immediately apparent, causing it to become open along $k_y$, consistent with having traversed a van Hove singularity at M$_2$.

In Fig. 2 we show two example data sets, from the extremes of strain reached in the experiment. For an unstrained sample mounted on a conventional sample plate (Fig. 2b), the three known bulk bands of Sr$_2$RuO$_4$ are clearly seen, with no signs of surface states (see Methods). The large, nearly circular γ sheet closes around Γ as an electron pocket, in agreement with a large number of previous measurements [14,15,23,35]. The data shown in Fig. 2d are from a sample for which an anisotropic strain of $\varepsilon_{xx}-\varepsilon_{yy} = -0.7\pm0.1\%$ was achieved, as determined by optical characterisation (see Fig. S2 of the Supplementary Information). In sharp contrast to the unstrained case, the γ sheet is no longer a circle but an open sheet along the y axis. This is in agreement with the calculations of how the Fermi surface of Sr$_2$RuO$_4$ would look after traversal of its van Hove singularity (vHs) located at $(0,\pm\pi)$ [4,6,36].

Confirmation that the vHs has indeed been traversed comes from inspection of the dispersions measured along the Γ-M$_1$ and Γ-M$_2$ directions.[*] For Γ-M$_1$ (Fig. 3a) the Fermi surface crossings of both the β and γ sheets are clearly visible, but along Γ-M$_2$ (Fig. 3b) the top of the γ band lies below the Fermi level. As seen in Fig. 3c, the combined Γ- M$_2$ - X cut reveals that this band displays the basic topography of the simple zone edge vHs predicted by band theory and sketched in Fig. 3d: the dispersion rises along $\Gamma$-$M_2$, then flattens at the saddle point and then rises slightly along $M_2 - X$ before the data are cut off by the Fermi function.

---

[*] In the strained sample, we distinguish between the M points located at $(\pm\pi,0)$ and $(0, \pm\pi)$, denoting these as M$_1$ and M$_2$, respectively.

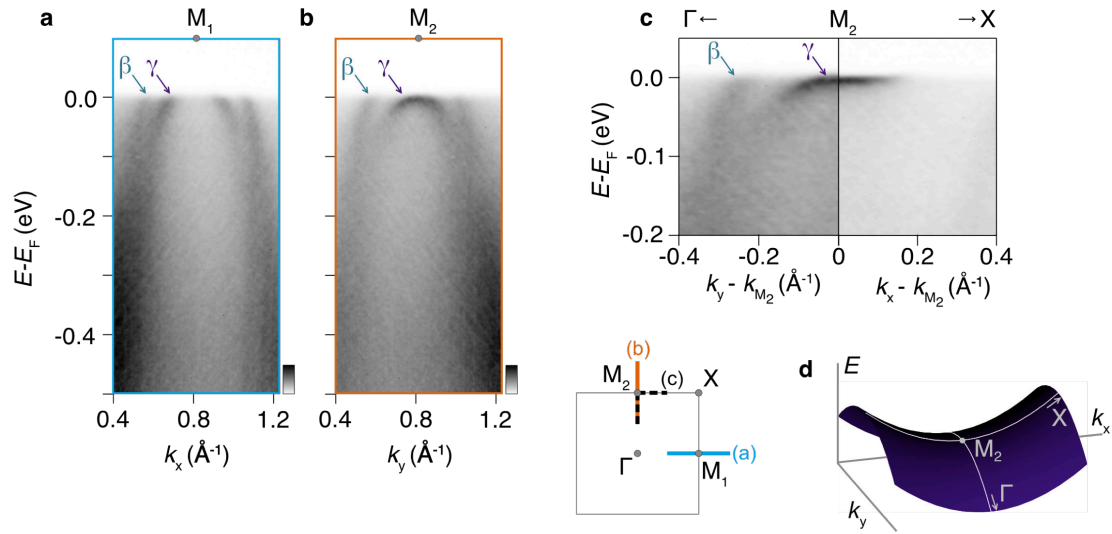

Fig. 3: **Strain-tuning to the van Hove singularity.** (a,b) Dispersions of $Sr_2RuO_4$ under anisotropic strain in the vicinity of the Brillouin zone boundary, measured along the (a) $\Gamma$-$M_1$ and (b) $\Gamma$-$M_2$ directions (see inset). The $\gamma$-band is clearly located above $E_F$ at the $M_1$ point, intersecting the Fermi level away from $M_1$ along the $\Gamma$-$M_1$ direction. In contrast, the $\gamma$-band is pushed below the Fermi level at $M_2$, with a fully-occupied parabolic band visible along $\Gamma$-$M_2$. (c) Measurements along the orthogonal directions away from $M_2$ (see inset) reveal the saddle point nature of the band dispersion at the M points, with a barely-occupied upward dispersing band visible along $M_2$-X before it is cut off by the Fermi function. The saddle point is shown schematically in (d).

The data in Figs. 2 and 3 firmly establish the qualitative result that we have been able to achieve a high enough uniaxial pressure to drive $Sr_2RuO_4$ through its Lifshitz transition at the $M_2$ point of the Brillouin zone. Moreover, the anisotropic strain for which we achieve this is in agreement within experimental error with that required[7,34] to reach a peak in the superconducting $T_c$, and at which the low-temperature resistivity deviates from a $T^2$ temperature-dependence (for details see Supplementary Fig. S2). This therefore provides compelling evidence that both are directly driven by tuning of the $\gamma$-band vHS to the Fermi level, a scenario also supported by analysis of the superconducting critical field [4], and NMR Knight shift [26] data.

It is desirable to track the strain-evolution of the Fermi surface approaching this van Hove singularity. Although our strain device based on thermal contraction is not inherently tuneable, it is in fact possible to achieve a range of sample strains by varying the sample thickness (see Methods). It is thus highly beneficial to have an internal measure of the strain achieved in every sample. Analysis of the $\beta$ sheet provides such a metric. The band topography makes its distortion much smaller than that of the $\gamma$ sheet; indeed it is hardly visible simply by looking at Fig. 2d. However, it exists, and can be traced by fitting momentum distribution curves extracted radially around the Fermi surface (dots in Fig. 4a). As shown in Fig. 4b, this analysis reveals how the β sheet $k_F$ along the Γ- $M_1$ and Γ- $M_2$ directions differ by 0.025 Å$^{-1}$ in the highly strained sample (Fig. 2d, reproduced with fits in Fig. 4a). This difference corresponds to an asymmetry, ($k_F$(Γ-$M_2$) - $k_F$(Γ-$M_1$))/($k_F$(Γ-$M_2$) +$k_F$(Γ-$M_1$)), of approximately 2%. This small change means that the β sheet distortion is likely in the linear response regime to a good approximation, so that the measured anisotropy can be used as a linear scale of the microscopic strain in every sample. The much larger γ sheet anisotropy is shown in an equivalent plot in Fig. 4c.

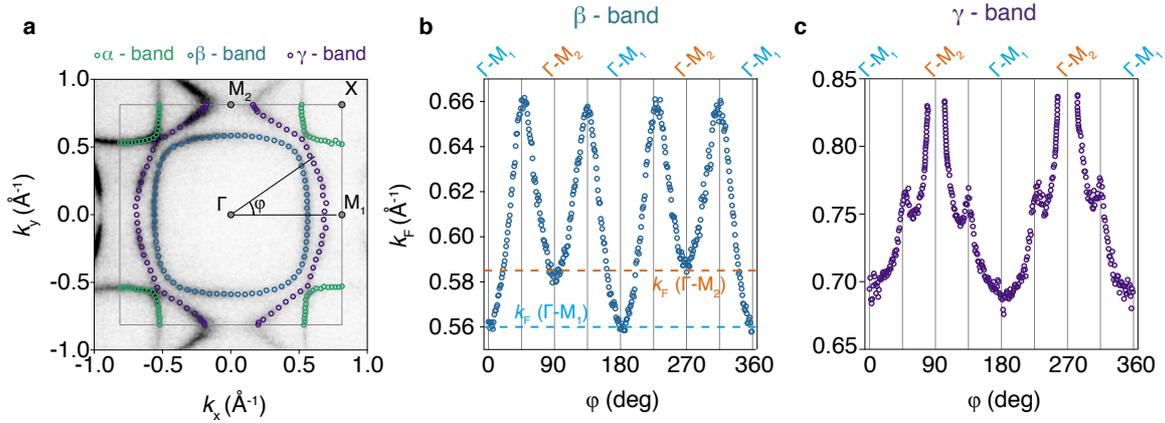

**Fig. 4: Quantitative analysis of Fermi surface anisotropy.** (a) Fermi surface of uniaxially-pressured Sr$_2$RuO$_4$ (image plot) with Fermi momenta (points) extracted from fitting MDCs. (b,c) Corresponding $k_F$ of the (b) β- and (c) γ-band Fermi surfaces as a function of radial angle, φ (see (a)). A small C$_2$ distortion of the β-sheet Fermi surface is evident, while the γ-sheet develops open contours.

Making use of the above-described β-band asymmetry, we show in Fig. 5a the strain dependence of the γ sheet M point anisotropy for five samples subjected to varying uniaxial stress (see Supplementary Fig. S3), including one pressurized with our original spring-based rig (see Supplementary Fig. S1). We parameterise the γ sheet distortions via the momentum separation of the γ Fermi surfaces in neighbouring Brillouin zones or, when its Fermi contour becomes open, by the momentum separation between the two branches along the Brillouin zone boundary. We define the latter as negative, to reflect its distinct topology. At $M_1$, Δ$k_F$ grows monotonically with increasing strain, reflecting an upwards shift of the vHs at this point, and hence reduction in $k_F$ of the γ-barrel along Γ-$M_1$. This is driven by the greater overlap of $d_{xy}$ orbitals along this compressively strained direction. Along Γ-$M_2$, a tensile strain is induced due to the positive Poisson's ratio of the sample platform, and the γ-band bandwidth consequently narrows, causing the vHs to drop below $E_F$ along this direction. Δ$k_F$ therefore reduces, and changes sign as the vHs is traversed.

This behaviour is qualitatively reproduced in the Fermi surface topology as calculated by DFT (Fig. 5b,c). However, to investigate whether this single-particle calculation correctly captures the strain dependent Fermi surface evolution, a more quantitative comparison is needed. Here we again make use of the β-band asymmetry as an internal reference of the anisotropic strain. Indeed, our calculations (Supplementary Fig. S4a) confirm that the β-band asymmetry is linearly proportional to the asymmetric strain. Moreover, they show that this metric is independent of the Poisson's ratio used in the calculation, providing an elegant way to compare results for samples mounted on platforms with freestanding samples, as, for example, investigated in Refs. [3,4,7,34].

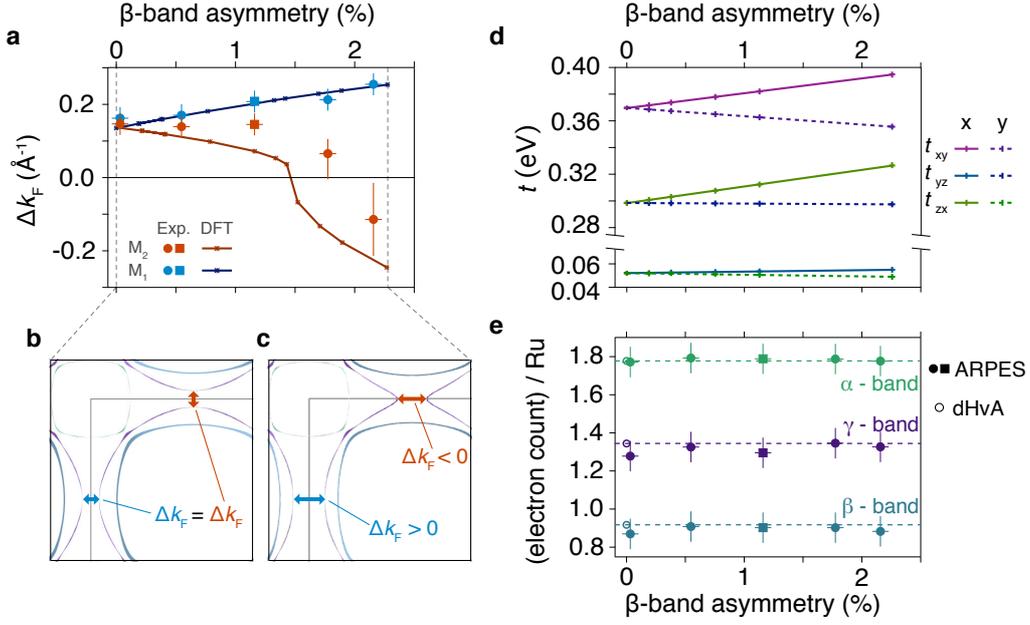

**Fig. 5: Strain-evolution of Fermi surface anisotropy.** (a) Parametrization of the γ-sheet anisotropy, encoded via the momentum separation of γ Fermi surfaces in neighboring Brillouin zones ($\Delta k_F>0$) or by the momentum separation between the two branches along the Brillouin zone boundary once the Fermi contour becomes open ($\Delta k_F<0$). These are plotted as a function of anisotropic strain encoded via the β-sheet asymmetry. The square symbols are from a sample mounted on the spring-based rig described in the Supplementary Information (Fig. S1). The measured Fermi surfaces and dispersions from all of the samples included here are shown in Supplementary Fig. S3. (b,c) Calculated Fermi surfaces for unstrained (b) and strained (c) $Sr_2RuO_4$. The line thickness encodes the degree of out-of-plane dispersion. (d) Downfolding the calculations onto a Wannier basis, we find that the hopping terms, shown here for nearest neighbours, vary linearly with strain with orbital-dependent prefactors. (e) Strain-dependent measurements of the Luttinger counts of the three Fermi surfaces, showing no resolvable changes as a function of strain.

Our calculations provide an excellent match to the measured strain-evolution of the γ Fermi surface in the vicinity of the $M_1$ point (Fig. 5a). Close to $M_2$, however, the agreement becomes much poorer as the Lifshitz transition is approached. The discrepancy between the DFT and the experiment is larger than the uncertainties associated with either (see also Supplementary Fig. S4b), likely reflecting a many-body contribution (see also Luo et al.[26]). At the single particle level, our DFT calculations downfolded on a Wannier tight-binding basis (see Methods) predict a linear scaling of the hopping parameters with strain (Fig. 5d), with orbital-dependent pre-factors. It is an interesting open question whether the discrepancies between the strain evolution of the Fermi surface predicted by such linear scaling and that found in our measurements can be understood on the basis of local self-energies, as in unstrained $Sr_2RuO_4$ [15], or may in fact imply that correlations become momentum-dependent in the vicinity of the Lifshitz transition.

Our findings therefore motivate future theoretical work studying the strain evolution of electronic correlations in $Sr_2RuO_4$, as well as providing important constraints for such studies. For example, we find that the Luttinger counts of each of the $\alpha$-, $\beta$- and $\gamma$-band Fermi surfaces (Fig. 5e) is, within our experimental uncertainty, independent of strain and consistent with the values known from de Haas van Alphen measurements in unstrained $Sr_2RuO_4$ [13]. This is in contrast to the case of biaxial epitaxial strain [23], for which approaching the vHs in $Sr_2RuO_4$ relies on a redistribution of charge carrier density between the $\alpha$, $\beta$ and $\gamma$ bands. The data in Fig. 5e indicate that uniaxial pressure tuning to the van Hove singularity instead results essentially entirely from distortion of the $\gamma$ band.

The results presented in this paper represent the first *k*-resolved spectroscopic evidence for the uniaxial stress-driven changes in the electronic structure of $Sr_2RuO_4$. Within experimental error, the strain at which ARPES shows that the van Hove singularity in the $\gamma$ sheet is reached is the same as the narrow range of strains at which there are strong peaks in $T_c$, the normal state NMR Knight shift and normal state resistivity. Our findings therefore provide strong evidence that, as previously postulated but not proven, all of these phenomena are associated with the Lifshitz transition caused by traversing this van Hove singularity. This has important implications for understanding the normal state and superconducting physics of $Sr_2RuO_4$, and offers the prospect of testing modern theories of its electronic structure and superconducting instability. We also believe that the basic passive platform design that enabled the experiments presented here will prove to be useful in other extreme environments, further establishing uniaxial pressure as a novel means of achieving disorder-free tuning of quantum materials.

**Methods**
**Differential thermal contraction strain rig:** We describe here the design and operation of our uniaxial stress apparatus shown in Fig. 1. This stage uses the differential thermal contraction between aluminium and titanium to uniaxially compress a sample platform. Aluminium contracts by 0.42% between room temperature and the measurement temperature, and titanium by 0.15%. This differential contraction is applied over a length of 6 mm, producing a thermal displacement of 16 μm. By necking the sample platform, its spring constant can be kept low relative to those of the other components, so that the resulting elastic deformation is concentrated into the neck. The spring constant of the two platforms together (there is a mirror of the sample platform on the bottom, to keep the device symmetric) is ≈8 N/μm. That of the remaining parts of the device, meaning the aluminium struts and titanium bars that generate the thermal displacement, is ≈20 N/μm, so 20/(20+8) ≈ 70% of the thermal displacement goes into the platforms. This compresses them uniaxially by ~0.6% between room temperature and below ~40 K, resulting in an anisotropic strain in the platform of $|\varepsilon_{xx}-\varepsilon_{yy}| \approx 0.8\%$. This value is confirmed by comparison of optical micrographs taken at room temperature and 10K (Supplementary Fig. S2). Crucially, this strain is applied gradually as the sample is cooled: the elastic limit of single-crystal $Sr_2RuO_4$ is as low as 0.15% at room temperature but at least 1% at 5 K [34], and by making use of differential thermal contraction in this way strain is applied to the sample as its elastic limit increases with cooling.

The sample is affixed to the necked portion of this platform using silver epoxy; this is the conventional sample mounting approach for ARPES measurements. Although this stage is not intrinsically tunable, the strain achieved in the sample varies with sample thickness, allowing different strains to be realised from different cleaves. This was most likely achieved through a combination of elastic and nonelastic deformation of the epoxy. The samples were in a size range permitting, even with fully elastic epoxy deformation, partial strain transmission. The datasheet for Epotek H21D silver epoxy indicates a room-temperature storage modulus, equivalent to the Young's modulus for elastic materials, of 5.5 GPa, while the Young's modulus for stress along a Ru-Ru bond direction in $Sr_2RuO_4$ is ~176 GPa [37]. This large difference in Young's moduli means that the strain in the sample locks to that in the platform over a length scale $\lambda$ that increases as the sample is made thicker. For epoxy and sample thicknesses both on the order of 10 μm, this length scale $\lambda$~100 μm. Our samples were typically ~600 μm across, larger but not drastically larger than $\lambda$, allowing meaningful variation in the achieved sample strain through varying sample thickness. Epoxy creep at higher temperatures is likely to have provided an additional mechanism to relax strain [38], which would also be more effective for thicker samples. Samples were cleaved at room temperature, so for cooling the sample stage from the epoxy curing temperature of 120°C to room temperature the samples were thicker – generally 50-150 μm thick – and so also mechanically stronger than during cooling from room temperature.

**Angle-resolved photoemission:** High-quality single-crystal $Sr_2RuO_4$ samples were grown in a floating zone furnace (Canon Machinery) using techniques refined over many years to those described recently in ref. [39]. These were cut into square platelets of dimensions ca. 600x600 μm² with the square edge oriented along [100] and, except where stated, were mounted on the custom sample stage shown in Fig. 1, with the [100] direction aligned to the uniaxial compression direction of the strain cell. The samples had varying thickness down to ca. 15 μm (Supplementary Fig. S2), enabling different strains to be achieved as discussed above. ARPES measurements were performed at the I05 beamline of Diamond Light Source [40], at a manipulator temperature of ~10 K. We used 68 eV linear horizontal (LH, *p*-polarised) photons for Fermi surface maps, 40 eV LH photons for measurements of the Γ-M dispersions, and 40 eV linear vertical (*s*-pol) light for measurement of the M-X dispersion, all chosen to ensure the most favourable transition matrix elements.

*In-situ* cleaved $Sr_2RuO_4$ is known to support surface states which substantially complicate the measured spectra in unstrained $Sr_2RuO_4$ [14]. It would be difficult to separate the strain-induced changes of the bulk electronic structure from the surface contributions. We therefore cleaved our samples in air immediately prior to loading them into the vacuum chamber. The resulting ARPES measurements of an unstrained reference sample mounted on a standard sample plate (Fig. 2(a)) reveal only the three well-known bulk Fermi surfaces of $Sr_2RuO_4$ [9], with no observable trace of surface-derived features. We therefore proceeded with this method for all of our measurements of strained $Sr_2RuO_4$.

**Density-functional theory:** Relativistic density functional (DFT) electronic structure calculations were performed using the full-potential local orbital FPLO code[41–43], version fplo18.00-52. For the exchange-correlation potential, within the local density approximation (LDA) the parametrizations of Perdew-Wang [44] was chosen. The spin-orbit (SO) coupling was treated non-perturbatively solving the four component Kohn-Sham-Dirac equation [45]. To obtain precise band structure and Fermi surface information, the final calculations were carried out on a well-converged mesh of 64.000 *k*-points (40x40x40 mesh, 8631 points in the irreducible wedge of the Brillouin zone). As starting point, for the unstrained crystal structure the structural parameters from Ref. [46] at 15 K have been used. Except where stated, the room-temperature experimental Poisson ratio was used for the calculations at finite strain [37] with the free internal structural parameters optimized, minimizing forces below 1meV/Å. A three band tight binding model was constructed from Ru centered Wannier functions for the 4d *xy*, *xz* and *yz* orbitals.

**Acknowledgements**
We thank T. Kim and M. Watson for useful discussions, and U. Nitzsche (IFW Dresden) for support with computational facilities. We gratefully acknowledge support from the European Research Council (Grant No. ERC-714193-QUESTDO), the Royal Society, the Max-Planck Society and the International Max-Planck Partnership for Measurement and Observation at the Quantum Limit. V.S. acknowledges EPSRC for PhD studentship support through grant number EP/L015110/1. E.A.M. and I.M. acknowledge PhD studentship support from the IMPRS for the Chemistry and Physics of Quantum Materials. N.K. acknowledges the support from JSPS KAKENHI (Nos. JP17H06136 and JP18K04715) and JST-Mirai Program (No. JPMJMI18A3) in Japan. We thank Diamond Light Source for access to beamline I05 (Proposal No. SI20427), which contributed to the results presented here.

# Supplementary Information: Direct Observation of a Uniaxial Stress-driven Lifshitz Transition in $Sr_2RuO_4$

V. Sunko[1,2], E. Abarca Morales[1,2], I. Marković[1,2], M.E. Barber[1], D. Milosavljević[1], F. Mazzola[2], D.A. Sokolov[1], N. Kikugawa[3], C. Cacho[4], P. Dudin[4], H. Rosner[1], C.W. Hicks[1,^], P.D.C. King[2,^] and A.P. Mackenzie[1,2,^]

[1] Max Planck Institute for Chemical Physics of Solids, Nöthnitzer Strasse 40, 01187 Dresden, Germany
[2] SUPA, School of Physics and Astronomy, University of St. Andrews, St. Andrews KY16 9SS, United Kingdom
[3] National Institute for Materials Science, Tsukuba, Ibaraki 305-0003, Japan.
[4] Diamond Light Source, Harwell Campus, Didcot, OX11 0DE, United Kingdom.
[^] to whom correspondence should be addressed: Clifford.Hicks@cpfs.mpg.de; philip.king@st-andrews.ac.uk; Andy.Mackenzie@cpfs.mpg.de

## 1) Spring-based rig

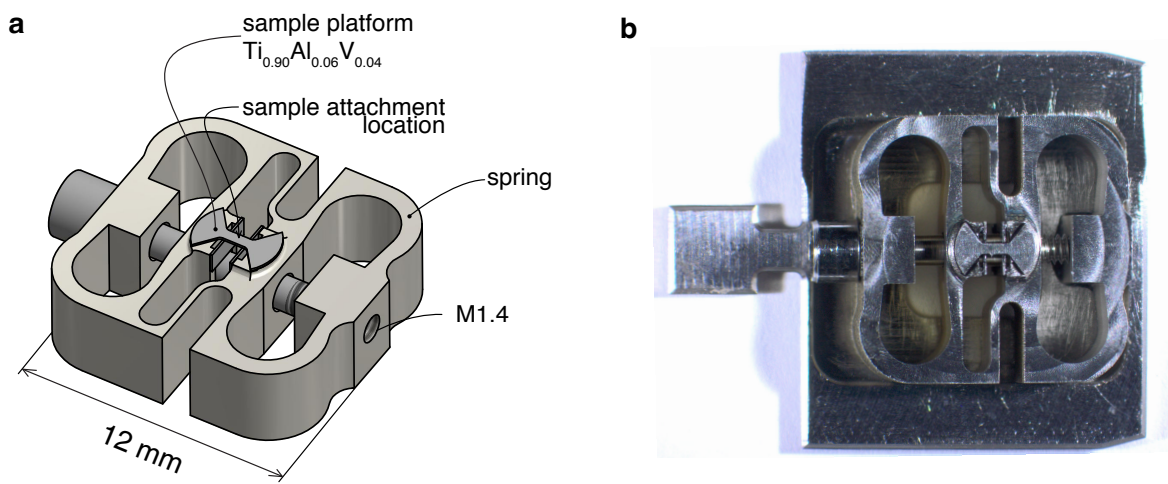

**Fig. S1: The spring-based rig.** (a) An illustration of the spring-based rig, actuated by *ex-situ* turning of the adjustment screw. The spring constant of the spring (0.8 N/μm) is designed to be 10 times smaller than that of the sample substrate. A displacement caused by turning the M1.4 screw by half a turn (0.15 mm) results in a sample platform strain of ~1%, as confirmed by optical analysis of the strained platform. Both the spring and the substrate were made of Grade 5 titanium ($Ti_{0.90}Al_{0.06}V_{0.04}$). (b) A photograph of the spring-based rig mounted on a standard flag-style sample plate.

## 2) Sample and platform characterisation

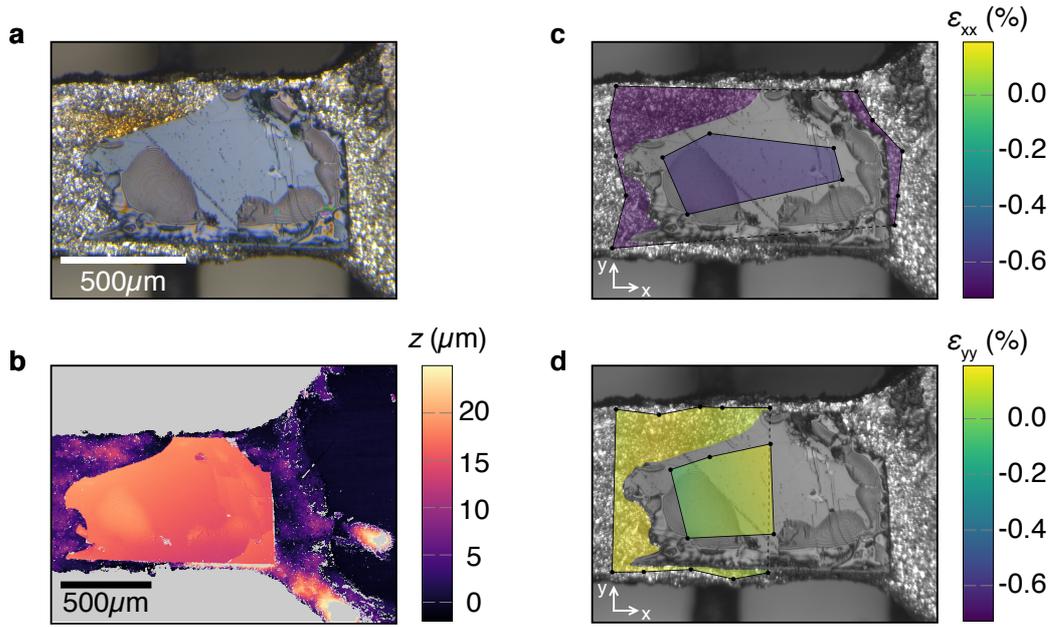

**Fig. S2: Platform and sample strain determination.** (a) An optical micrograph of the maximally-strained sample, the data from which are shown in Figs. 2–4, taken after the ARPES measurements. (b) The thickness of the sample and epoxy measured with an optical profilometer. (c)-(d) The strain achieved in the sample and platform was determined by tracking the relative displacement of features on their surfaces (indicated by the dots) as the device was cooled down in an optical cryostat. The average strain developed by cooling from room temperature to ~10 K was found to be $\varepsilon_{xx}$ = -0.61±0.03% and $\varepsilon_{yy}$ = +0.05±0.10% in the sample, and $\varepsilon_{xx}$ = -0.71±0.02% and $\varepsilon_{yy}$ = +0.12±0.07% in the platform. These values correspond to anisotropic strains of $\varepsilon_{xx}-\varepsilon_{yy}$ = -0.7±0.1% in the sample, and -0.8±0.1% in the platform, as quoted in the main text.

The best current estimate of the strain value at which the peak in $T_c$ and associated anomalies in the electronic properties of $Sr_2RuO_4$ are observed is obtained by measurements on free-beam samples using a calibrated force sensor[1]. They showed that the peak in low-temperature resistivity is reached for a uniaxial pressure of 0.7 GPa, a value that can be converted to an anisotropic strain of $\varepsilon_{xx}-\varepsilon_{yy}$ = -0.55% using the known room-temperature values for Young's modulus and Poisson's ratio (176 GPa and 0.39, respectively [2]). The $T_c$ peaks at a strain ~10% higher than the low-temperature resistivity [3], i.e. at an anisotropic strain of $\varepsilon_{xx}-\varepsilon_{yy}$ = -0.61%.

## 3) Strain-dependent ARPES data

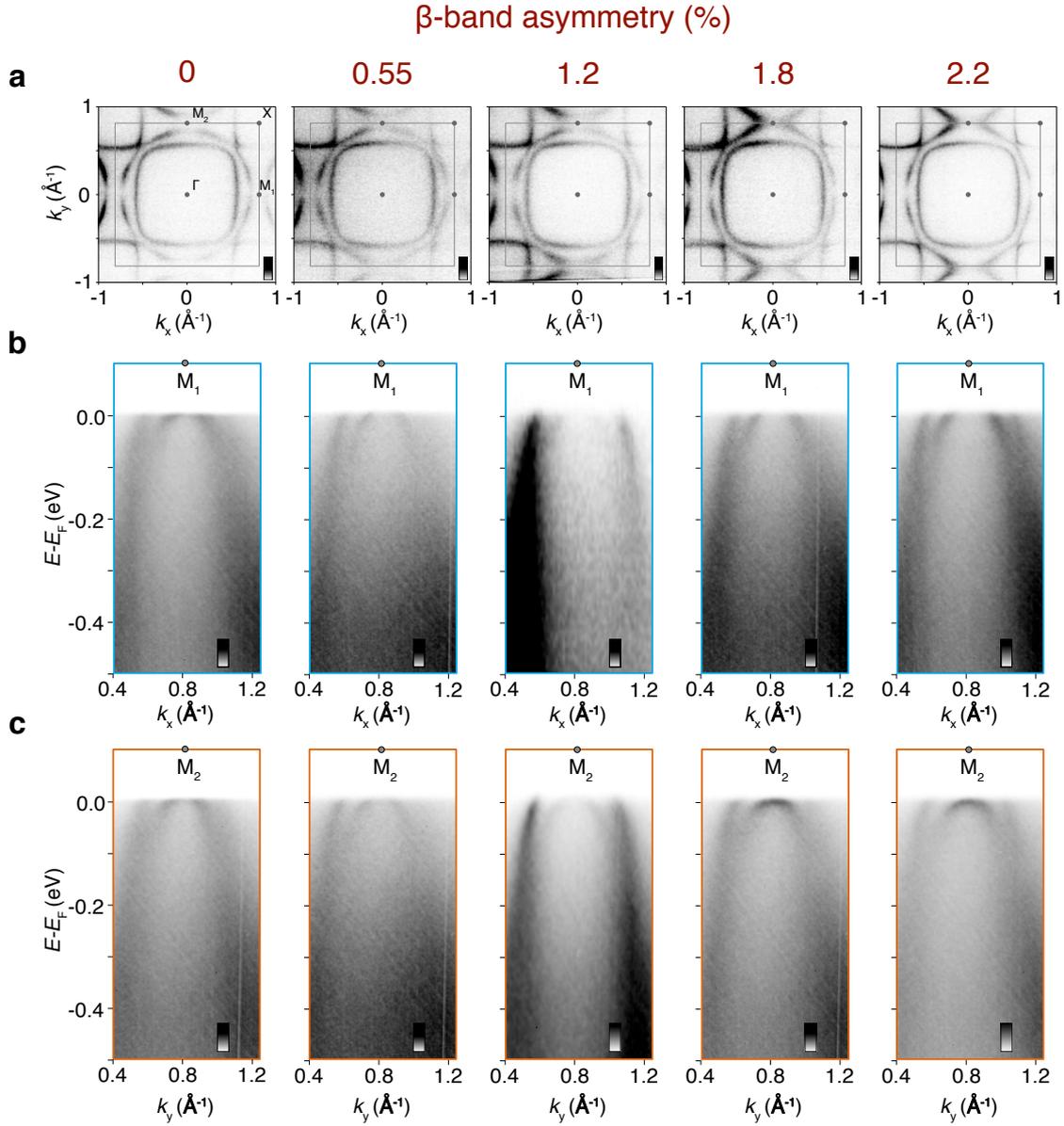

**Fig. S3: Strain-dependent ARPES measurements.** Measured (a) Fermi surfaces and (b,c) dispersions along the (b) Γ- $M_1$ and (c) Γ- $M_2$ direction from samples under different strain, as encoded by their varying β-band asymmetries. The measurements of the unstrained sample (β-band asymmetry of 0%) were taken on a standard sample plate, while the measurements at the β-band asymmetry of 1.2% were taken on a sample mounted on the spring-based cell (Fig. S1). All other samples were mounted on the differential thermal contraction sample stage (Fig. 1 of the main text). All the Fermi surface maps were measured using a photon energy of 68eV, as were the dispersions at the β-band asymmetry of 1.2%. All other dispersions were measured using a photon energy of 40eV. All measurements were taken using *p*-polarised light. The points in Fig. 5(a,c) of the main text were extracted from fitting of these measured dispersions and Fermi surfaces, respectively.

## 4) Insensitivity of calculated Fermi surface anisotropies on Poisson's ratio

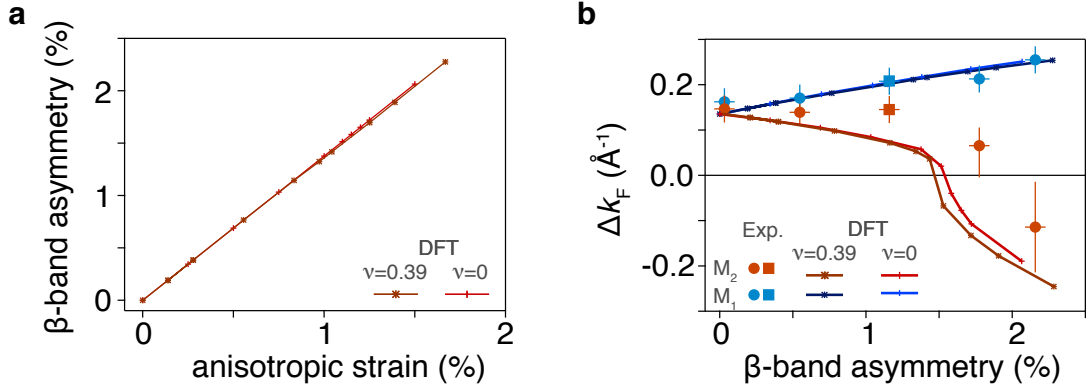

**Fig. S4: Insensitivity of calculated Fermi surface anisotropies on Poisson's ratio.** (a) The β-band asymmetry as a function of the absolute value of the anisotropic strain, $|\varepsilon_{xx}-\varepsilon_{yy}|$, as calculated by DFT for $Sr_2RuO_4$ with its experimental Poisson's ratio of $\nu = 0.39$, and assuming a pure uniaxial strain ($\nu = 0$). The calculated β-band asymmetry is linear with anisotropic strain, with a slope independent of Poisson's ratio, confirming β-band asymmetry as a useful internal metric of anisotropic strain. (b) Parametrization of the γ-sheet anisotropy, as a function of uniaxial strain encoded via the β-sheet anisotropy (same as Figure 5a of the main text). DFT calculations are performed using the experimental Poisson's ratio of $\nu = 0.39$, as well as assuming a pure uniaxial strain ($\nu = 0$). The good agreement between the two calculations shows that the discrepancy between DFT and experiment cannot be accounted for by the fact the sample on the platform expands according to the Poisson ratio of the platform, rather than the sample.